\documentclass[sigconf]{acmart}
\usepackage{popets}

\setcopyright{popets}
\copyrightyear{2024}

\acmYear{2024}
\acmVolume{2024}
\acmNumber{3}
\acmISBN{}
\acmConference{Proceedings on Privacy Enhancing Technologies}
\usepackage[scaled]{beramono}
\usepackage[T1]{fontenc}
\usepackage[english]{babel}
\usepackage{blindtext}
\usepackage{natbib}
\usepackage{microtype}
\usepackage{multirow}
\usepackage{booktabs}
\usepackage{xspace}
\usepackage[labelformat=simple]{subfig}
\usepackage{pgf}
\usepackage{adjustbox}
\usepackage{tcolorbox}
\usepackage{algorithm}
\usepackage{algpseudocode}
\usepackage{balance}
\usepackage[bookmarksnumbered,unicode]{hyperref}
\hypersetup{hidelinks}
\usepackage{amsfonts}
\usepackage[frozencache]{minted}
\usepackage{enumitem}
\usepackage[font=small,labelfont=bf]{caption}

\definecolor{cbone}  {HTML}{005B94} 
\definecolor{cbtwo}  {HTML}{FF800E} 
\definecolor{cbthree}{HTML}{ABABAB} 
\definecolor{cbfour} {HTML}{595959} 
\definecolor{cbfive} {HTML}{5F9ED1} 
\definecolor{cbsix}  {HTML}{C85200} 
\definecolor{cbseven}{HTML}{898989} 
\definecolor{cbeight}{HTML}{A2C8EC} 
\definecolor{cbnine} {HTML}{FFBC79} 
\definecolor{cbten}  {HTML}{CFCFCF} 

\colorlet{old}{cbone}
\colorlet{primarycolor}{cbone}
\colorlet{primaryshaded}{cbfive}
\colorlet{secondarycolor}{cbtwo}
\colorlet{secondaryshaded}{cbnine}
\colorlet{tertiarycolor}{cbfour}
\colorlet{tertiaryshaded}{cbten}

\usepackage{tikz}
\usepackage{pgfplots}
\pgfplotsset{compat=1.13}
\usetikzlibrary{calc}
\usetikzlibrary{positioning,arrows,fit,matrix}
\usetikzlibrary{backgrounds}
\usetikzlibrary{patterns}
\usetikzlibrary{shapes.arrows} 
\usepgfplotslibrary{groupplots}

\usepackage{listings}
\lstdefinestyle{mydefaultcode}
{
	numbers=left,
	framexleftmargin=1pt, 
	numbersep=10pt, 
}
\newcommand{\colcode}[2]{\colorbox{primarycolor!15}{\makebox(#2,4){\strut#1}}}
\newcommand{\culcode}[2]{\colorbox{secondarycolor!15}{\makebox(#2,4){\strut#1}}}
\newcommand{\czlcode}[2]{\colorbox{tertiarycolor!15}{\makebox(#2,4){\strut#1}}}
\newcommand{\ccpp}[3]{\colorbox{primarycolor!#1}{\makebox(#3,4){\strut#2}}}
\newcommand{\ccps}[3]{\colorbox{primarycolor!#1}{\makebox(#3,4){\strut#2}}}
\newcommand{\ccpt}[3]{\colorbox{primarycolor!#1}{\makebox(#3,4){\strut#2}}}
\newcommand{\ccpg}[3]{\colorbox{primarycolor!#1}{\makebox(#3,4){\strut#2}}}

\usepackage{amsmath}
\usepackage{bm}
\newtheorem{defin}{Definition}

\newtheorem{thm}{Theorem}
\DeclareMathOperator*{\argmax}{argmax}

\usepackage[capitalize,noabbrev,nameinlink]{cleveref}
\crefname{lem}{Lemma}{Lemmas}
\crefname{defin}{Definition}{Definitions}
\crefname{thm}{Theorem}{Theorems}
\usepackage{makecell}

\newcommand{\appref}[1]{\hyperref[#1]{Appendix~\ref{#1}}}

\newcommand{\caliskan}{\citeauthor{IslHarLiuNar+15}\xspace}
\newcommand{\abuhamad}{\citeauthor{AbuAbuMohNya+18}\xspace}
\newcommand{\code}[1]{{\small\color{primarycolor!90!black}\texttt{#1}}}
\newcommand{\codeintext}[1]{\mbox{\color{primarycolor!90!black}\textproc{#1}}}

\newcommand{\eg}{e.\,g.}

\makeatletter
\renewcommand{\verbatim@font}{\small\ttfamily}
\makeatother

\begin{document}

\renewcommand{\sectionautorefname}{Section} 
\renewcommand{\subsectionautorefname}{Section} 
\newcommand{\algorithmautorefname}{Algorithm} 
\newcommand{\subfigureautorefname}{\figureautorefname} 
\renewcommand{\thesubfigure}{(\alph{subfigure})} 

\newcommand{\prg}{p}
\newcommand{\prgs}{P}
\newcommand{\attr}{\mathcal{A}}
\newcommand{\anonym}{\mathcal{Y}}
\newcommand{\uncer}{u}
\newcommand{\near}{N_{t,k}}
\newcommand{\acceptedrule}{\rule[2.5pt]{4.5cm}{0.2pt}}
\newcommand{\acceptedbegin}{\normalfont \small \itshape \acceptedrule}

\title[On Challenges in Anonymizing Source Code]{
  {\acceptedbegin Accepted at the Privacy Enhancing Technologies Symposium (PETS) 2023 \acceptedrule}\\
  I still know it's you!\\On Challenges in Anonymizing Source Code
}

\author{Micha Horlboge}
\affiliation{
  \institution{Technische Universität Berlin}
  \country{}
}

\author{Erwin Quiring}
\affiliation{
  \institution{ICSI \& Ruhr Universität Bochum}
  \country{}
}

\author{Roland Meyer}
\affiliation{
  \institution{Technische Universität Braunschweig}
  \country{}
}

\author{Konrad Rieck}
\affiliation{
  \institution{BIFOLD \& Technische Universität Berlin}
  \country{}
}

\begin{abstract}
  The source code of a program not only defines its semantics but also
  contains subtle clues that can identify its author. Several studies
  have shown that these clues can be automatically extracted using
  machine learning and allow for determining a program's author among
  hundreds of programmers. This attribution poses a significant threat
  to developers of anti-censorship and privacy-enhancing technologies,
  as they become identifiable and may be prosecuted. An ideal
  protection from this threat would be the \emph{anonymization of
    source code}. However, neither theoretical nor practical
  principles of such an anonymization have been explored so far.
  In this paper, we tackle this problem and develop a framework for
  reasoning about code anonymization. We prove that the task of
  generating a \mbox{\emph{$k$-anonymous program}}---a program that
  cannot be attributed to one of $k$ authors---is not computable in the
  general case.
  As a remedy, we introduce a relaxed concept called
  \emph{$k$-uncertainty}, which enables us to measure the protection
  of developers. Based on this concept, we empirically study candidate
  techniques for anonymization, such as code normalization, coding
  style imitation, and code obfuscation. We find that none of the
  techniques provides sufficient protection when the attacker is aware
  of the anonymization.
  While we observe a notable reduction in attribution performance on
  real-world code, a reliable protection is not achieved for all
  developers. We conclude that code anonymization is a hard problem
  that requires further attention from the research community.
\end{abstract}


\maketitle

\section{Introduction}
\label{sec:introduction}

The source code of a program provides a wealth of information for
analysis. It not only defines syntax and semantics, but also contains
clues suitable for identifying its author. These clues result from the
personal \emph{coding style} and range from obvious programming
habits, such as the naming of variables and functions, to subtle
preferences in the usage of data types, control structures, and
API~\citep{SimZetKoh2018}. Thus, similar to writing style in
literature, a source code unnoticeably carries a fingerprint of its
developer. Several studies have shown that this coding style can be
automatically extracted using machine learning and allows for
identifying the author of a program among hundreds of other
developers~\citep[e.g.,][]{KalKauGon+20, BogKovRebBac2021,
  AlrShiWanDeb+18, IslHarLiuNar+15}. As an example,
\citet{AbuAbuMohNya+18} report a detection accuracy of 96\% on a
dataset of source code from 1,600 developers participating in a coding
competition. Although these methods still struggle under realistic
conditions, some approaches reach up to 61\% on only fragments of
real-world code from 104 programmers~\citep{DauCalHarShe2019}.

While authorship attribution of source code resembles a valuable tool
for digital forensics, it also poses a threat to developers of
anti-censorship and privacy-enhancing technologies. Anonymous
contributors to open-source projects, such as Tor~\citep{DinMatSyv04}
and I2P~\cite{NguKinAntPol18}, become identifiable through
learning-based attribution and might be prosecuted for their work in
repressive countries. Unfortunately, defenses against this threat have
received little attention so far.  Even worse, prior work has shown
that strong obfuscation of source code is still not sufficient to
prevent an attribution~\citep[see][]{IslHarLiuNar+15, AbuAbuMohNya+18,
  CalYamDauHar+18}, indicating the challenge of protecting developers.

In this paper, we tackle this problem and study the
\emph{\mbox{anonymization} of source code} from a theoretical and
practical perspective. To this end, we propose a framework for
reasoning about code anonymization and attribution. Based on this
framework, we introduce the concept of a \emph{$k$-anonymous} program,
that is, a program that cannot be attributed to one of $k$ authors and
hence is protected by an anonymity set. We prove that changing a given
source code, so that it becomes $k$-anonymous in the general case is
unfortunately not computable and resembles an undecidable problem.
Consequently, a universal method for code anonymization cannot exist
and so the search of practical protection is a challenge for research.

As a remedy, we derive a relaxed concept that we denote as
\mbox{\emph{$k$-uncertainty}}. Instead of a program being perfectly
indistinguishable between authors, we require that it is attributed to
$k$~authors with similar confidence. While this concept cannot overcome
the undecidability of $k$-anonymity, it provides a novel means for
\emph{measuring} the protection of developers. By inspecting the
confidence range of the $k$ most similar authors in an attribution, we
can evaluate how well a developer is hidden in an anonymity set. Based
on this concept, we introduce a numerical measure called
\emph{uncertainty score} that ranges from 0 (no protection) to
1 ($k$-anonymity) and can be used to empirically assess how well a
source code is protected. As a result, it becomes possible to
empirically compare techniques for protecting the identity of
developers.

Based on this numerical measure, we conduct a series of experiments to
analyze candidate techniques for code anonymization. In particular, we
consider \emph{code normalization},
\emph{coding style imitation}~\citep{QuiMaiRie19} and
\mbox{\emph{code obfuscation}~\citep{tigress,stunnix}} as defenses
against popular attribution methods
\citep{IslHarLiuNar+15, AbuAbuMohNya+18}. For our experiments, we work
with two datasets, one from a programming competition with 30
developers and the other from open-source projects of 81 developers.
At first, all techniques hinder an attribution and lead to high
uncertainty scores. However, their performances diminishes once the
attacker becomes aware of the protection. For the strongest technique,
the popular obfuscator Tigress~\citep{tigress}, the attribution still
reaches an accuracy up to 24\% and 8\%, respectively. To understand
this result, we develop a method for explaining the attributions and
uncover clues in the source code that remain after anonymization. This
explanation method complements our uncertainty score by indicating weak
spots in the realized protection.

When iteratively removing clues with our method from the competition
dataset, we eventually bring the source code to an uncertainty score
close to 1. However, this result should not be interpreted as a defeat
of the attribution methods. Rather, it shows that anonymization can be
achieved in a limited and controlled setup. The systematic removal of clues, however, cannot simply be transferred to the real world, where
neither the attribution method nor the learning data is known to a
developer. We thus argue that there is a need for novel anonymization
concepts and consider our work as the first step towards formalizing
and evaluating these approaches.
In summary, we make the following contributions:
\begin{itemize}[leftmargin=18pt]
  \setlength{\itemsep}{3pt}
  \item \emph{Theoretical view on code anonymization.} We propose a
        framework for reasoning about code anonymization. This
        enables us to prove that universal $k$-anonymity cannot be reached.

  \item \emph{Practical view on code anonymization.} We introduce a
        concept for measuring anonymization. Based on this,
        we empirically compare protection techniques under different
        adversaries.

  \item \emph{Insights on obstacles of code anonymization.} Finally, we
        develop an approach for explaining attribution methods and
        identifying clues remaining after an anonymization
        attempt.
\end{itemize}

\noindent\textbf{Roadmap.} We review authorship attribution of source
code in \autoref{sec:background} and discuss our threat model in
\autoref{sec:motivation}. Our framework for analyzing code anonymity
is introduced in \autoref{sec:code-anonymity} and we empirically
evaluate different techniques with it in
\autoref{sec:anonym-under-test}. We analyze the deficits of the
techniques in \autoref{sec:impr-code-anonym}. Limitations and related
work are presented in \autoref{sec:disc--limit} and
\autoref{sec:related-work}, respectively. \autoref{sec:conclusion}
concludes the paper.

\section{Source Code Attribution}
\label{sec:background}

We start with a short primer on authorship attribution of source
code. The objective of this task is to automatically attribute a given
source code to its author based on individual properties of coding
style~\citep{KalKauGon+20}. As these properties are hard to formally
characterize, this objective is typically achieved by extracting
\emph{features} from source code and constructing an attribution method
using \emph{machine learning}. Existing approaches are hence best
described based on the considered features and the employed learning
algorithms.

\subsection{Features of Source Code}

The features currently used in authorship attribution roughly fall into
three types: layout, lexical, and syntactic. Each of them relies on a
different representation and thus provides different types of stylistic
patterns. We briefly review these features in the following.

\subsubsection{Layout Features}

The first type of features are derived from the layout of the source
code. Developers often have specific preferences to format their code,
such as different forms of indentation.
\autoref{fig:imitator-code-example-layout} shows a function, where
different features are highlighted. Even in this short snippet a
variety of layout features is visible, such as the indentation width
of 4. Consequently, any attempt to anonymize code needs to start by
removing individual formatting.

\begin{figure}[t]
  \centering

\begin{tikzpicture}[
	arrowcode/.style = {thick, 
		>=stealth',}, 
	annotationstyle/.style={
		align=left,anchor=west, font=\small,
	},
	arrowcustomstyle/.style={
		->, arrowcode, shorten <= 0.1cm
	},
	layoutarrow/.style={arrowcustomstyle, dashdotted, primarycolor},
	lexicalarrow/.style={arrowcustomstyle, dashed, secondarycolor},
	syntacticarrow/.style={arrowcustomstyle, dotted, tertiarycolor},
	]
	
\node[] (code) at (0,0) {
\begin{minipage}[t]{0.55\textwidth}
\begin{lstlisting}[
	style=mydefaultcode,
%	frame=tblr, % draw a frame around code block; we dot it below!
	escapeinside={\§§}{*)},    
	language=C++,
	linewidth=6.0cm,
	xleftmargin=.10\columnwidth, % necessary for centering
	xrightmargin=.10\columnwidth, % necessary for centering
	]
§§\culcode{int}{8}*) gcd(§§\culcode{int}{8}*) a, §§\culcode{int}{8}*) b) {
§§\colcode{    }{12}*)if(b == 0)
§§\colcode{    }{12}*)§§\hspace{1pt}*)§§\colcode{    }{12}*)return a;
§§\colcode{    }{12}*)return gcd(b, §§\czlcode{a}{1}*) % §§\czlcode{b}{1}*));
}

\end{lstlisting}
\end{minipage}
};

\node[draw, minimum width=4.30cm, minimum height=2.10cm] at 
(.455,0.025) {};

\def\rights {3.00};
\node at (\rights, 0) (rightsnode) {};
\def\mid {0.45};
\def\midwithcomments {2.1};

\node[annotationstyle] (C1) at (\rights, 1.05) 
{\textit{\# Token int = 3}\\\textit{(Lexical)}};
\draw[lexicalarrow] (\rights, 1.25) -| (-1.29, 0.88);
\draw[lexicalarrow] (\rights, 1.25) -| (-0.14, 0.88);
\draw[lexicalarrow] (\rights, 1.25) -| ( 0.88, 0.88);

\node[annotationstyle] (C2) at (\rights,0.17) 
{\textit{Indentation width is 4}\\\textit{(Layout)}};
\draw[layoutarrow] (\rights, 0.36) -- (0.60, 0.36);
%

\draw[syntacticarrow] (\rights, -0.83) -| (1.08, -0.55);
\draw[syntacticarrow] (\rights, -0.83) -| (1.74, -0.55);
\node[annotationstyle] (C4) at (\rights,  -0.70) {\textit{Leaves in 
AST}\\\textit{(Syntactic)}};

\end{tikzpicture}
  \vspace{-6pt}
  \caption{Code snippet in C with highlighted feature types.}
  \label{fig:imitator-code-example-layout}
  \vspace{-4pt}
\end{figure}

\subsubsection{Lexical Features}

The second type of features is derived from the lexical analysis of
source code. The resulting features comprise identifiers, keywords,
literals, operators, and other terminal symbols of the underlying
grammar~\citep{AhoSetUll06}. These features implicitly encode the
syntax and semantics of the source code. For example,
\autoref{fig:imitator-code-example-layout} shows a lexical feature
that counts the occurrence of the token \code{int}. This reflects
a developer's preference for this type in relation to others, such as
\code{long} or \code{int32\_t}. Compared to the layout, lexical
features cannot be unified easily, as they are implicitly linked to
syntax and semantics.

\subsubsection{Syntactic Features}

Finally, the syntax of source code provides further features for
characterizing the programming habits of developers. In particular,
the \emph{abstract syntax tree} (AST) is a common representation that
allows extracting patterns in the types, arithmetics, logic, and
control flow used by developers~\citep{IslHarLiuNar+15,
  AlsDauHarMan+17, BogKovRebBac2021}. These features range from single
language constructs to syntactic fragments, such as tree-like
structures in the AST.  \autoref{fig:imitator-code-example-layout}
highlights two code locations that correspond to leaves in the AST.
Syntactic features are hard to modify. Replacing a single keyword in
the source code may lead to several modifications in the AST.
Similarly, adapting one node of the tree may require multiple code
modifications. The removal of coding style thus becomes challenging, as
we demonstrate in \autoref{sec:anonym-under-test}.

\subsection{Attribution Using Machine Learning}

The described features provide a complex view on source code that is
difficult to interpret by a human analyst. State-of-the-art attribution
methods for source code therefore rely on machine learning to recognize
stylistic patterns for a particular author~\mbox{\cite[e.g.,][]{Pel00,
    AbuAbuMohNya+18, IslHarLiuNar+15, AlsDauHarMan+17}}. To this end, a
supervised learning algorithm is applied to infer characteristics for
each author. The result is a \emph{multiclass classifier} that returns
confidences for all authors from the training data. The highest-ranked
author is typically selected for attribution.

Previous work has studied several learning algorithms for this
attribution, such as support vector machines~\cite{Pel00}, random
forests~\cite{IslHarLiuNar+15}, and deep neural
networks~\cite{AlsDauHarMan+17, AbuAbuMohNya+18}. As we find in our
evaluation, the algorithm can have a considerable impact on
anonymization. For example, deep neural networks tend to overfit to
particular authors, while other algorithms provide a more generalized
prediction.

\section{Motivation and Threat Model}
\label{sec:motivation}

Code authorship attribution is typically considered an instrument of
forensics, similar to stylometry~\mbox{\citep[e.g.,][]{DinFunIqbChe2019,
    TerPasBlaSua2022, AfrIslStoGre+14}}. For example,
\citet{CalYamDauHar+18} show how binary code can be attributed to
malware authors, aiding the prosecution of cybercrime. Unfortunately,
however, authorship attribution can become a malicious tool itself when
used by repressive countries to pursue the developers of
regime-critical software, such as anti-censorship and anonymization
tools. While there are no reports of automated attribution yet, a
recent incident in China underscores the pressure on developers: On
November 2, 2023, over 20 developers, some of which operate under a
pseudonym, simultaneously removed their anti-censorship software from
public repositories, likely indicating a concerted action against them
\citep{china-pet1,china-pet2}. Similar incidents have already occurred
in the past~\citep{china-pet3,china-pet4}, albeit not to this extent.

Any programmer contributing code to open-source code under their real
name runs the risk that software developed later under a pseudonym will
be linked to their identity. This risk increases if the prosecutor can
narrow down a group of individuals and thus simplify the attribution
task. In our evaluation in \autoref{sec:anonym-under-test}, we show
that this risk is fortunately lower than in the common benchmarks for
authorship attribution. Nevertheless, every third developer is
correctly identified in our main experiment, raising the question of how
such de-anonymization can be prevented, what techniques are available,
and whether they offer sufficient protection.

\subsection{Defending against Attribution}

At a first glance, the anonymization of software may seem like a
straightforward task: The developer needs to manipulate their code such
that the attribution method is tricked into predicting the wrong
author, similar to an \emph{adversarial example}~\mbox{
  \citep{SimZetKoh2018, QuiMaiRie19, MatStaPrePer+19}}. However, there
is no reason for a repressive regime to focus only on the most likely
person indicated by an attribution method; other highly ranked
individuals can also be prosecuted. If we do not know where the
prosecutor stops their investigation of the attribution ranking,
defending becomes hard. In this case, protection is not about deceiving
the attribution but ensuring that one person's coding style is
indistinguishable from that of others.

The two plots in \autoref{fig:illustrate} illustrate this setting.
While an adversarial example can easily cause a misclassification by
crossing the decision boundary of an attribution method, it does not
provide reliable protection. The true author is still identifiable due
to the large differences in the attribution confidences. By contrast,
in the right plot, the software is moved near to the intersection of
the decision boundary so that several of the developers become
similarly likely. The author is protected from identification by an
anonymity set. In this work, we set out to investigate this form of
\emph{code anonymity} and analyze its technical feasibility.

\begin{figure}[h]
  \centering
  \vspace{-4pt}
  \subfloat[Misclassification\label{fig:illustrate-mis}]{
    \scalebox{.99}{\begin{tikzpicture}[framed,scale=1.0,font=\footnotesize,
	markstyle/.style={only marks, mark size=2.2},
	arrow/.style = {shorten <= 4pt, shorten >= 4pt, thick, 
		>=stealth',}
	]

	\input{./tikz/anonymity_misclassification_base.tex}
	
	\draw[secondarycolor] plot [markstyle, mark=square*] 
	coordinates 
	{
		(-1.1, -0.2)
	};
	
	\draw[->,  arrow] (-1.1, 0.8) to[out=225, in=135] 
	(-1.1, -0.2);
	
	\node[anchor=north east,xshift=-2pt] at (-1.1, -0.2) 
	{$\overline{\prg}$};
	
	\node[align=center] at (0.2,1.8) (anon) 
	{\phantom{Anonymity Set}};

	\begin{axis}[at={(-0.15\linewidth,-0.33\linewidth)},
		height=0.3\columnwidth,
		width=0.50\columnwidth,
		tick align=outside,
		enlarge x limits=0.20,
				ymin=0,
		ymax=0.8,
		major tick length = 2pt,
		yticklabels={,,},
		ylabel={$c$},
		ylabel shift=-2pt,
		title={Confidences for $\overline{\prg}$},
		title style={yshift=-1.5ex},
		every axis plot/.append style={
			ybar,
			bar width=.65,
			bar shift=0pt,
			fill,draw
		},
        xtick={1,2,3,4},
		xticklabels={A,B,C,D},
		]
	\addplot[secondarycolor]coordinates {(1,0.6)};
	\addplot[tertiarycolor]coordinates{(2,0.7)};
	\addplot[tertiarycolor]coordinates{(3,0.1)};
	\addplot[tertiarycolor]coordinates{(4,0.1)};

	\end{axis}

\end{tikzpicture}}
  }
  \quad
  \subfloat[Anonymization\label{fig:illustrate-anon}]{
    \scalebox{.99}{\begin{tikzpicture}[framed,scale=1.0,font=\footnotesize,
	markstyle/.style={only marks, mark size=2.2},
	arrow/.style = {shorten <= 4pt, shorten >= 4pt, thick, 
		>=stealth',}
	]

	\draw[tertiarycolor, thick, dashed, fill=tertiaryshaded!20] (0,0) 
	circle (16pt);

	\input{./tikz/anonymity_misclassification_base.tex}

	\draw[secondarycolor] plot [markstyle, mark=square*, 
	mark 
	size=2.5] 
	coordinates 
	{
		(0,0)
	};
	\draw[->,  arrow] (-1.1, 0.8) to[out=250, in=200]
	node[pos=0.5,anchor=north east] {$\anonym$}
	(0,0);
	
	\node[anchor=north west,xshift=2pt,yshift=2pt] at (0,0) 
	{$\tilde{\prg}$};
		
	\node[align=center] at (0.2,1.8) (anon) 
	{Anonymity Set};
	\draw[tertiarycolor] (-0.02, 0.33) -- (anon); 

	\begin{axis}[at={(-0.15\linewidth,-0.33\linewidth)},
	height=0.3\columnwidth,
	width=0.50\columnwidth,
	tick align=outside,
	enlarge x limits=0.20,
	ymin=0,
	ymax=0.8,
	major tick length = 2pt,
	yticklabels={,,},
	ylabel={$c$},
	ylabel shift=-2pt,
	title={Confidences for $\tilde{\prg}$},
	title style={yshift=-1.5ex},
	every axis plot/.append style={
		ybar,
		bar width=.65,
		bar shift=0pt,
		fill,draw
	},
	xtick={1,2,3,4},
	xticklabels={A,B,C,D},
	]
	\addplot[secondarycolor]coordinates {(1,0.2)};
	\addplot[tertiarycolor]coordinates{(2,0.2)};
	\addplot[tertiarycolor]coordinates{(3,0.2)};
	\addplot[tertiarycolor]coordinates{(4,0.2)};
	
\end{axis}		
\end{tikzpicture}}
  }
  \caption{Schematic comparison of misclassification and
    anonymization in code authorship attribution.}
  \label{fig:illustrate}
  \vspace{-3pt}
\end{figure}

\subsection{Threat Model}
\label{sec:threat-model}

To provide a basis for investigating code anonymity, we define a
threat model consisting of an \emph{attacker} and a \emph{defender}.

\subsubsection{Attacker}
The attacker analyzes software with unknown authorship. Their goal is
to determine whether some of the code has been developed by the
defender. The attacker knows a group of suspected developers and thus
operates in a closed-world setup. In addition, the attacker has access
to training data, that is, software developed by the suspects under
their real identity, for instance, in open-source or work projects.
This enables the attacker to apply machine learning and compute a
ranking of likely suspects based on the confidence of the learning
model.

\subsubsection{Defender}
The defender aims to avoid being identified when publishing specific
software, such as anti-censorship tools. The defender has no knowledge
of the attribution method employed by attacker. Similarly, the defender
does not know the ranking or number of likely suspects. However, they
can arbitrarily change their code as long as its semantics are
preserved. When modifying their software, the defender prefers changes
that preserve readability, so strong obfuscation is only considered as
a last resort.

\section{Code Anonymity}
\label{sec:code-anonymity}

To the best of our knowledge, there exists no previous work exploring
code anonymization. Hence, we first introduce a unified notation for
describing programs (software) and their semantics.

\subsection{Unified Notation}

We denote a \emph{program} by $\prg \in \prgs$ where $\prgs$ is the set
of all valid programs. We differentiate between the
\emph{representation} and \emph{semantics} of a program $\prg$, where
the former defines its code, such as the source code, while the latter
describes its behavior~\citep{AhoSetUll06}. If two programs $\prg_a$
and $\prg_b$ have the same representation, that is, the source code is
identical, we write $\prg_a = \prg_b$.  If two programs implement the
same semantics, that is, their behavior and output is identical for all
inputs, we write $\prg_a \equiv \prg_b$.

This differentiation enables us to investigate the relation of
representation and semantics: If we have $\prg_a = \prg_b$, it directly
follows that $\prg_a \equiv \prg_b$. The opposite, however, does not
hold. Rich programming languages, like C and C++, enable implementing
the same behavior in infinite many ways. For example, identifiers can
be changed and API functions can be substituted. Given a program
$\prg_a$, there typically exist many $\prg_b \in \prgs$, such that
$\prg_a \equiv \prg_b$ but $\prg_a \neq \prg_b$. As an example,
\autoref{fig:examples} in the appendix shows four programs that are
semantically equivalent yet make use of different identifiers, types,
control flow and API functions. This asymmetry between representation
and semantics fuels the hope that anonymizing code might be a
relatively simple task.

\subsubsection{Anonymization and Attribution}
We continue to introduce notation for attribution and anonymization
methods. In particular, to identify the author of a given program, we
define a generic \emph{attribution method}
\begin{equation}
  \attr : \prgs \rightarrow (0,1)^n, ~~ \prg \mapsto c = (c_1, ..., c_n)
  \label{eq:attr}
\end{equation}
that maps a program $\prg$ to a vector $c$ of $n$ values
$c_1, ..., c_n$, each associated with the confidence for one of $n$
possible authors. Without loss of generality, we assume that $\attr$ is
deterministic and attains a performance at least as good as random
guessing.

In practice, attribution methods typically return the author associated
with maximum confidence, that is, $\argmax \attr(\prg)$. However, all
current approaches for learning-based attribution provide a measure of
confidence, such as the class probabilities returned by a random forest
or a neural network \citep{IslHarLiuNar+15, AbuAbuMohNya+18,
  AlsDauHarMan+17}. Consequently, they all fit our generic definition of
$\attr$. As an example, in \autoref{fig:illustrate-mis} the attribution
method $\attr$ returns the confidence vector
$\attr(\bar{\prg}) = (0.6,0.7,0.1,0.1)$.

As antagonist to the attribution method in our analysis, we introduce a
generic \emph{anonymization method}
\begin{equation}
  \anonym : \prgs \rightarrow \prgs, ~~ \prg \mapsto \tilde{\prg}
  ~~\text{with}~~\prg \equiv \tilde{\prg}
  \label{eq:anon}
\end{equation}
where the anonymized program $\tilde{\prg}$ is semantically equivalent
to $\prg$ but possess properties that obstruct the attribution. Without
loss of generality, we assume that the anonymization remains in the set
$\prgs$ of valid programs.  For example, $\prgs$ could be defined as
all programs that solve a particular task, and thus any
semantic-preserving transformation remains within this set. In
\autoref{fig:illustrate-anon}, the anonymization method $\anonym$
changes the attribution, so that we have
$\attr(\anonym(\prg)) = (0.25,0.25,0.25,0.25)$.

Note that we do not explicitly model the feature types and learning
algorithms within $\attr$ or the code transformations performed by
$\anonym$ at this point. In a our threat model, the defender is not
aware of the employed attribution method and thus an analysis of the
underlying feature space and the impact of code transformations on the
features cannot be anticipated.

\subsection{Modeling Anonymity}
\label{sec:modeling-anonymity}

Equipped with this notation, we are ready to formally model the
anonymity of source code. For this purpose, we build on the classic
concept of \emph{$k$-anonymity} proposed by \mbox{\citet{Swe02}} and
expand it to authorship attribution. As we see in the following, even
this simple concept with known weaknesses is hard to realize on
programs.

\begin{defin}[k-anonymity]
  \label{def:kanon}
  Given an attribution method $\attr$, a program $\prg$ is
  $k$-anonymous, if the attribution confidence $c_t$ of the true author
  is identical to the confidence values of at least $k-1$ other
  authors. That is, for $\attr(\prg) = c$ holds
  $c_t = c_i = \ldots = c_{i+k-1}$ and $\argmax \attr(p)$ is not unique.
\end{defin}

For ease of presentation, we reference the $k-1$ authors in sequential
order from $i$ to $i+k-1$, although their indices may be arbitrary
distributed in the vector $c$.  This definition implies that for a
$k$-anonymous program, the true author is indistinguishable from at
least $k-1$ other authors and thus remains hidden in an anonymity set
of size $k$.  Hence, an anonymization method realizing $k$-anonymity
transforms a given program, so that it resides at the exact
intersection of the classification function for $k$ authors in the
feature space, as shown in \autoref{fig:illustrate}.

The above definition is not directly applicable in our threat model, as
the defender has not knowledge of the attribution method $\attr$.
Hence, we introduce \emph{universal $k$-anonymity}, an extended
definition which aims to protect against any possible attribution
method, thus compensating the missing knowledge of the defender.

\begin{defin}[Universal k-anonymity]
  \label{def:ukanon}
  A program $\prg$ is universal $k$-anonymous, if it is $k$-anonymous
  for any possible attribution method $\attr$.
\end{defin}

Although \cref{def:ukanon} may seem like a good start for reasoning
about attribution and designing methods for code anonymization, it
already reaches the general limits of computability.

\begin{thm}
  \label{thm:undecide}
  Given a program $\prg$, the problem of transforming $\prg$ using an
  anonymization method $\anonym$ so that $\anonym(\prg)$ is universal
  \mbox{$k$-anonymous} is incomputable (undecidable).
\end{thm}

\begin{proof}
  We reduce the problem of \emph{program equivalence}, which is known
  to be undecidable~\citep{GolJac12}, to the task of creating universal
  $k$-anonymity. Let $\prg_a$ and $\prg_b$ be two programs written by
  developers $a$ and $b$ with $\prg_a \neq \prg_b$. Furthermore, let
  $\anonym$ be an anonymization method whose output is universal
  $k$-anonymous. Then, the programs are semantically equivalent if and
  only if their anonymization yields the same representation, that is,
  \begin{equation}
    \anonym(\prg_a) = \anonym(\prg_b) ~~ \Longleftrightarrow ~~ \prg_a \equiv \prg_b.
  \end{equation}

  To understand this reduction, let us suppose the programs are
  semantically equivalent. Then, as long as $\anonym(\prg_a)$ and
  $\anonym(\prg_b)$ are not identical, there always exists an
  attribution method $\attr_\delta$ that can differentiate the
  developers. This $\attr_\delta$ can be constructed as follows: We
  describe the difference between the anonymized programs as
  $\delta = \anonym(\prg_a) \, \backslash \, \anonym(\prg_b)$, where we
  assume that $\delta \neq \varnothing$. As the programs are
  semantically equivalent, the difference $\delta$ can only result from
  the coding style of the developers. Thus, we can define
  $\attr_\delta$ as
  \begin{equation}
    \attr_\delta(\prg) =
    \begin{cases}
      (1, 0) & \text{if $\delta$ is in $\prg$}, \\
      (0, 1) & \text{otherwise}.
    \end{cases}
  \end{equation}
  Since $\attr_\delta$ can be constructed for any difference $\delta$,
  the method $\anonym$ is forced to normalize the programs to the same
  representation, such that we have $\anonym(\prg_a) = \anonym(\prg_b)$.
  If the programs are not semantically equivalent, their anonymized
  representation can never be identical and we always get
  $\anonym(\prg_a) \neq \anonym(\prg_b)$.  As a result, a method
  $\anonym$ creating universal $k$-anonymous programs would solve the
  undecidable problem of program equivalence and thus is incomputable.
\end{proof}

\cref{thm:undecide} fundamentally limits our ability to anonymize code.
Although $k$-anonymity is a rather weak concept that suffers from
well-known shortcomings \cite{MacKifGehVen07,LiLiVen07}, we are not
even able to establish it on source code when the attribution method is
unknown. In view of the great flexibility of expressing semantics in
code, this is a surprising, negative result that unveils the challenges
of protecting developers from identification.

\smallskip
\begin{tcolorbox}[boxrule=0.75pt,colback=white,colframe=cbone,sharp corners=all]
  \emph{Takeaway message.~} The problem of creating universal
  \mbox{$k$-anonymity} on source code is incomputable. Although
  theoretically appealing, the development of approaches to solve this
  problem for Turing-complete programming languages is a dead end for
  research.
\end{tcolorbox}
\medskip

\subsection{Relaxing Anonymity}

As a consequence of this situation, we lift our requirements and seek a
weaker definition of code anonymity. To this end, we propose a relaxed
form of an anonymity set: Instead of requiring $k$~authors to receive
an identical attribution, we demand that their confidence values lie
close to each other, that is, within an interval of a small value
$\epsilon$. An anonymization method now needs to transform a program so
that it is close to the intersection of the classification function,
yet it is not forced to create identical programs. This relaxation is
illustrated in the right plot of \autoref{fig:illustrate} where the
vicinity of the intersection is indicated by a circle.

To model this concept, we consider the $k$-nearest neighbors of an
author $t$ in the confidence vector $c$. In particular, we define a
permutation $\pi$ of $c$ that sorts the confidences according to their
distance from $c_t$ in ascending order. The $k$-nearest neighbors can
then be defined as a sequence $\near$ as follows
\begin{align}
  \near = ( c_{\pi[1]}, c_{\pi[2]}, \ldots, c_{\pi[k]} )
\end{align}
where the true author's confidence is the first element and we always
have $|c_{\pi[i]} - c_t| \leq |c_{\pi[j]} - c_t| $ for any $i < j$.
Based on this relaxed form of an anonymity set, we introduce a new
concept for anonymity that we denote as \emph{$k$-uncertainty}. This
concept is a generalization of $k$-anonymity from \cref{def:kanon},
where for $\epsilon = 0$, both concepts are equivalent.

\begin{defin}[k-Uncertainty]
  \label{def:kuncer}
  Given an attribution method~$\attr$, a program $\prg$ is
  $k$-uncertain, if there exist at least $k-1$ other authors whose
  confidence values are $\epsilon$-close to the true author. That is,
  for $\attr(\prg) = c$ holds $\max(\near) - \min(\near) \leq \epsilon$.
\end{defin}

Since $k$-uncertainty is a generalization of $k$-anonymity, it inherits
undecidability and is also incomputable when the attribution method is
unknown. However, instead of enforcing a binary notion of anonymization
($k$-anonymous or not), this concept introduces a continuous level of
anonymity ($0 \leq \epsilon \leq 1$). As we see in the following, this
representation enables us to construct a measure for assessing the
protection of developers in practice.

\subsection{Measuring $k$-Uncertainty}
\label{sec:measuring-anonymity}

The concept of $k$-uncertainty gives rise to a
\emph{quantitative measure} of code anonymization. Instead of fixing
$\epsilon$, we can determine the size of the interval around an
author's $k$-nearest neighbors and thus gauge how well a program can be
attributed to that author. To achieve this goal, we define a
corresponding \emph{uncertainty score}
\begin{equation}
  \uncer_k(t, c) =  1 - (\max(\near) - \min(\near))
\end{equation}
that takes a normalized confidence vector $c$ as input and returns the
attribution uncertainty for the author $t$ based on \cref{def:kuncer}.
If the author is clearly identifiable, this score returns 0, whereas if
she is perfectly hidden in an anonymity set, we obtain 1. As an
example, let us consider the confidence vector
$c = ( 0.8, 0.1, 0.1, 0.0 )$ with $k = 3$. We immediately see that the
first author stands out from the rest. This exposure is also reflected
in the uncertainty score $\uncer_3 (1, c) = 0.3$.  The second author
cannot be clearly separated from the nearest neighbors, yielding
$\uncer_3 (2, c) = 0.9$.

Note that this use of a threshold $\epsilon$ deviates from typical
privacy research, where $\epsilon$ is fixed in advance. By contrast, we
fill the other variables in \cref{def:kuncer} and use $\epsilon$ as the
output. This unusual inversion allows for empirical analysis of
existing protection and attribution methods. Since theoretical
guarantees are currently not in reach, we argue that this is the next
feasible step towards understanding and limiting the identification of
developers.

\subsection{Interpreting $k$-Uncertainty}
\label{sec:interpreting-uncert}

The uncertainty score provides us with a numerical measure for
anonymity, yet its interpretation is not straightforward. The score
depends on the particular type of confidence values. If these values
correspond to class probabilities, as in many learning algorithms, we
have $\sum_{i=1}^n c_i = 1$ and can thus define a heuristic for
determining a threshold $t_\epsilon$ on the value of $\epsilon$.

For class probabilities, the maximum confidence of an author needs to
be above $\frac{1}{n}$ to make a reliable attribution, as otherwise the
method would not be better than random guessing. Consequently, we
define $ t_\epsilon = \frac{1}{n} $.  This ensures that the $k$ authors
of the anonymity set lie within an interval that is smaller or equal to
the confidence of random guessing. With this heuristic, we can also
interpret the uncertainty score and reason that scores above
$1 - t_\epsilon$ provide practical \mbox{$k$-uncertainty} on class
probabilities. We must emphasize, however, that this heuristic is not
generally applicable and must be carefully considered for each type of
confidence values.

\subsection{Alternative Measures}
\label{sec:uscore-vs-acc}

As with any new measure, it is essential to question its necessity and
explore the use of existing measures instead. While we do not claim
that the proposed uncertainty score is the only way to quantitatively
assess anonymity, we argue that it offers advantages over other
performance measures.  To understand these differences, recall that we
focus on an attacker who is not limited to identifying the first
predicted author of a software. Instead, they can analyze confidence
values and use all information therein to find suspects to pursue,
such as the top-$k$ predicted authors or developers with suspicious
gaps in confidence.

As a result, classic measures for attribution performance, such as the
\emph{accuracy} and the \emph{F-measure}, are not suitable. They only
reflect whether a prediction is correct and ignore the remaining
information in the ranking and confidence. Whether the actual author is
right next to the top prediction or in the middle of the ranking does
not make a difference. Consequently, these measures overestimate the
protection and may flag an anonymization approach as secure, although
the defender is identifiable with little extra effort.

Performance measures from information retrieval partially address this
problem~\citep{ManSchRag08}. For example, measures such as
\emph{mean reciprocal rank} and \emph{top-$k$-accuracy} are
specifically designed to take into account the order of predictions and
thus where the true author ranks. Nonetheless, examining the ranking
alone does not provide sufficient protection: First, the defender
cannot anticipate the number of top $k$ entries investigated by the
attacker. Second, unusual gaps in the confidence values may still
indicate suspicious authors, regardless of their ranking. Even worse,
if the attacker knew the defender hides below rank $k$, they could
adapt their strategy accordingly, weakening the protection again.

We argue that a quantitative measure of protection must include a
notion of privacy and not only attribution performance. That is, the
actual author should not only be ranked low but also protected within
an anonymity set of developers with similar confidence, making
identification much more difficult. It is precisely this idea that
forms the concept of our uncertainty score. We demonstrate in
\autoref{sec:anonym-under-test} that this notion of privacy pays
off and helps to evaluate protection techniques more accurately.

\section{Anonymization under Test}
\label{sec:anonym-under-test}

Prepared with a practical definition of anonymity, we can now take a
look at different approaches for protecting developers. Our goal is to
put these approaches to the test and assess how well they can realize
$k$-uncertainty in different scenarios. In particular, we study a
\emph{static scenario}, where the adversary is unaware of the
anonymization, and an \emph{adaptive scenario}, where she adapts the
attribution to it. Before presenting these tests, we introduce the
\emph{candidate techniques} for anonymization and our
\emph{evaluation setup}.

\subsection{Candidate Techniques}
\label{sec:cand-anonym-}

As there exist no approaches explicitly designed for anonymizing source
code, we focus on techniques that reduce the presence of coding style.
Specifically, we examine three candidate techniques:
\emph{code normalization}, \emph{coding style imitation}, and
\emph{code obfuscation}. All three differ in the amount and precision
of their modifications.

\subsubsection{Code Normalization}

The goal of normalization is to modify code so that it conforms to a
given policy or style guide. Normalization is regularly employed in
collaborative software development, and larger projects typically
define detailed guidelines for the layout and structure of code
\citep[e.g.,~][]{linux-style, chrome-style, firefox-style}. Inspired by
the available style guidelines, we develop a strong code normalization
and make it available to the research community. Our normalization
builds on 13 transformation rules for C code that unify the code
layout, replace the names of variables and functions, reduce the
variety of data types, and simplify control structures. All rules
preserve the program semantics, so that the normalization complies with
our definition of an anonymization method. \autoref{tab:norm-rules} in
\appref{sec:appendix-normalization} provides a detailed listing of the
implemented rules.

Note that code normalization can be applied without access to an
attribution method and thus provides a generic approach for reducing
the presence of stylistic patterns in source code.

\subsubsection{Coding Style Imitation}

As second candidate, we consider techniques that can imitate the coding
style of developers. In particular, we focus on approaches for creating
adversarial examples of source code~\citep{QuiMaiRie19,LiuJiLiuWu2021}.
In contrast to normalization, these attacks require access to an
attribution method and allow more target-oriented code modifications.
Typically, adversarial examples of source code are realized in a
two-stage procedure: First, a set of code transformations is defined,
each imitating a stylistic pattern, such as adding or removing
preferences for particular data types or control structures. Second,
these transformations are chained together using a search strategy
until a target classification is reached. This procedures also
preserves the semantics of the code.

There exist different variants for creating adversarial examples on
source code.  For our tests, we focus on the method by
\citet{QuiMaiRie19}, as it does not only induce misclassifications of
the attribution method, but also enables lowering its confidence. While
the objective of the method technically remains misclassification, we
conjecture that the low confidence better protects the author and thus
might serve as a suitable anonymization approach.

\subsubsection{Code Obfuscation}

As third technique, we consider the obfuscation of source code. This
candidate aims to make code incomprehensible to humans while preserving
its semantics. Technically, this can be achieved by, for example,
encrypting constants or obscuring control flow. We refer the reader to
the book by \mbox{\citet{NagCol09}} for an introduction to this topic.
Obfuscation is agnostic to the attribution method and can be employed
to any available code. For our experiments, we make use of two common
obfuscators, \emph{Stunnix}~\citep{stunnix} and
\emph{Tigress}~\citep{tigress}. Stunnix obfuscates identifiers,
constants and literals. Still, the overall structure of the program
remains unchanged. Tigress is a more sophisticated tool and considered
state of the art in obfuscation. It supports several advanced
obfuscation techniques, such as function merging and code
virtualization~\citep[see][for details]{tigress}.

Note that obfuscation is intended to prevent an understanding of code
and not its attribution. Hence, obfuscators only implicitly destroy the
coding style of developers.

\subsection{Evaluation Setup}
\label{sec:evaluation-setup}

Before testing the different candidate techniques, we introduce our
evaluation setup, which follows the common design of experiments with
code attribution~\citep{KalKauGon+20}.

\subsubsection{Evaluation Datasets}

As basis for our evaluation, we consider two datasets of source code in
the language C. We restrict our dataset to plain C and do not consider
C++, since the obfuscator Tigress only works with this language and
several features of C++ hinder code transformations, such as dynamic
bindings.

\smallskip
\emph{GCJ Dataset.}
The first dataset called \emph{GCJ} contains code written by 30 authors
as part of the \textit{Google Code Jam}~\citep{google-code-jam}
competition between 2012 and 2014.  All authors solved the same 8
tasks, so the differences in their solutions are caused by their
individual coding style. In total, our dataset contains 240~source
files. Similar datasets are commonly used to evaluate attribution
methods~\mbox{\citep[see][]{IslHarLiuNar+15, AbuAbuMohNya+18}}. For our
experiments, we use a \textit{grouped k-fold} split to select seven of
the eight problems for training and reserve the last one for testing.
Since the source code of this dataset comes from a competition, it is
not fully representative of real-world programming.

\smallskip
\emph{GH Dataset.}
The second dataset called \emph{GH} has been crawled from GitHub. It
contains source code in the C language written by 81 developers. The
code has been collected according to the procedure described in
\appref{sec:gith-crawl-proc} and contains 391 repositories with a total
of 1,284~source files. To simulate the threat model described in
\autoref{sec:threat-model}, we split the data along the
repositories into training and test sets. That is, one repository for
each author is considered unknown and used for testing, while the
others serve as training data. In contrast to the GCJ dataset, the
collected code comes from actual software projects and exhibits a wide
variety of functionality, complicating the task of inferring coding
style.

\smallskip
Before extracting features from both datasets, we expand all macros,
remove comments and use \textit{clang-format}~\citep{clang18} to
eliminate trivial layout differences. While this process requires
little effort for the GCJ dataset, we need to establish functional
build environments for several of the GitHub repositories, which is a
labor-intensive task, consuming about one person month of work.

\subsubsection{Attribution Methods}

We employ two state-of-the-art attribution methods to evaluate the
effectiveness of the candidate techniques: the method by
\citet{IslHarLiuNar+15} based on a random forest and the method by
\citet{AbuAbuMohNya+18} which primarily uses a recurrent neural
network. The approaches differ in the extracted features, where
\caliskan employ a mixture of lexical and syntactic features, while
\abuhamad use lexical tokens only. As a result, we gain insights on how
the learning algorithms and the features impact an anonymization.
Technically, we build on the framework of \citet{QuiMaiRie19} and the
corresponding implementations. As a coherence check, we compare the
performance of our setup with the original C++ dataset used by
\citeauthor{QuiMaiRie19} The results are shown in \autoref{tab:reprod}
and differ only slightly, indicating a valid experimental setup.

\begin{table}[h]
  \caption{Attribution performance (accuracy) as reported by
    \citet{QuiMaiRie19} and reproduced by us on C++ source code.}
  \label{tab:reprod}
  \vspace{3pt}
  \centering \small
  \vspace{-4pt}
  \begin{tabular}{lcc}
    \toprule
    \bf Attribution & \bf Quiring et al. & \bf Our implementation \\
    \midrule
    \caliskan       & 0.904 $\pm$ 0.02   & 0.901 $\pm$ 0.02       \\
    \abuhamad       & 0.884 $\pm$ 0.04   & 0.879 $\pm$ 0.05       \\
    \bottomrule
  \end{tabular}
\end{table}

When applying the methods of \caliskan and \abuhamad to the two
datasets of C source code considered for our evaluation, the
performance changes significantly, as shown in \autoref{tab:default}.
The accuracy decreases by up to 22\% points for the GCJ dataset and
62\% points for the GH dataset. We identify three factors contributing
to this drop: First, we remove all comments and layout features during
pre-processing, which eliminates trivial clues for discriminating
developers. Second, we focus only on C code, which is less diverse in
comparison to C++. Third, the GH dataset consists of real-world code,
which is characterized by a large variety of functionalities.
Consequently, less information about the coding style is accessible for
attribution. Nevertheless, roughly one out of three authors is
correctly attributed by the two methods for both datasets,
demonstrating the need for code anonymization.

\begin{table}[hbp]
  \caption{Attribution performance without any protection on the
    GCJ and GH datasets of C source code.}
  \label{tab:default}
  \vspace{3pt}
  \centering \small
  \vspace{-4pt}
  \begin{tabular}{llcc}
    \toprule
    \bf Data & \bf Attribution & \bf Accuracy & \bf Uncertainty Score \\
    \midrule
    GCJ      & \caliskan       & 0.688        & 0.840                 \\
    GCJ      & \abuhamad       & 0.754        & 0.261                 \\
    \midrule
    GH       & \caliskan       & 0.284        & 0.900                 \\
    GH       & \abuhamad       & 0.346        & 0.686                 \\
    \bottomrule
  \end{tabular}
\end{table}

In addition, \autoref{tab:default} shows our new uncertainty score for
the two attribution methods for $k=5$. Despite their similar
performance, the uncertainty scores differ considerably. The approach
of \caliskan yields 0.84 on the GCJ dataset, whereas the method of
\abuhamad reaches only 0.26, indicating large differences in confidence
between the authors. We examine these disparities later in Sections
\ref{sec:results-stat-attr} and \ref{sec:results-adapt-attr}. For
different values of $k$, the uncertainty score changes only slightly
and thus we keep $k=5$ for the remaining experiments.

\subsubsection{Candidate Techniques}

We implement the code normalization using \emph{LibTooling}, a
versatile library of the LLVM infrastructure~\citep{clang18}. For the
coding style imitation, we again build on the framework by
\mbox{\citet{QuiMaiRie19}} and fit it to our setup. For the code
obfuscation, we employ Stunnix in version 4.7 and Tigress in version
3.1. For Stunnix, we enable all options, while for Tigress we focus on
advanced features, such as virtualizing functions, inserting random
code, and obscuring API calls. \autoref{tab:tigress-transformations} in
\appref{sec:appendix-tigress-transformations} lists the used features.
We ensure that both tools are given random seeds so that randomized
elements are different in each run.

\begin{figure}
  \centering
  \vspace{-4pt}
  \input{./pgf/acc_static.pgf}
  \vspace{-18pt}
  \caption{Attribution performance (accuracy) of candidate techniques
    in the \emph{static} attribution scenario (regular training).}
  \label{tab:candidates}
\end{figure}

\subsection{Static Attribution Scenario}
\label{sec:results-stat-attr}

In our first scenario, we consider a \emph{static attribution}, where
the adversary is unaware of the employed anonymization techniques and
treats the modified code as regular programs. For this purpose, we
apply the considered techniques for anonymization to the
\emph{test set only} and investigate their impact on the accuracy and
uncertainty of the attribution methods. This setup reflects situations
where the attacker overlooks the presence of manipulated code, for
example, when the coding style is imitated.

\subsubsection{Attribution Performance}

\autoref{tab:candidates} shows the performance of the attribution
methods when the four candidate techniques are employed. We observe a
huge drop in accuracy compared to the original results in
\autoref{tab:default}. Obfuscation~I (Tigress) has the largest impact
and changes the code so that the accuracy decreases dramatically on
both datasets. The attained attribution of developers is no better than
random guessing. In contrast, Obfuscation~II (Stunnix) shows a weaker
protection and reduces the accuracy by only a few percentage points on
the GH dataset. The code normalization and coding style imitation
reduce the accuracy on the GCJ dataset. For the GH dataset, however,
only a minor reduction is observable for the normalization. The
imitation provides no protection, since it has been tailored towards
GCJ-style code \mbox{\citep[see][]{QuiMaiRie19}} and lacks sufficient
transformations to effectively manipulate the real-world code from the
open-source projects of the GH dataset.

\begin{figure}[t]
  \centering
  \vspace{-4pt}
  \input{./pgf/uncert_static.pgf}
  \vspace{-18pt}
  \caption{Anonymization performance (uncertainty score) in the
    \emph{static} attribution scenario (regular training).}
  \label{tab:uncertain}
\end{figure}

\subsubsection{Anonymization Performance}

We continue to investigate the anonymization performance of the
candidate techniques. \autoref{tab:uncertain} shows the average
uncertainty score with $k = 5$ for both datasets, that is, an anonymity
set of 5 authors. Compared to the original results, there is a
significant increase of this measure, indicating better protection of
the developers. For the method by \caliskan, all values are now above
0.91, while for the approach by \abuhamad all but three scores reach
over 0.79. The best performance is obtained for Tigress, reaching an
uncertainty score of over 0.97 for both attribution methods on both
datasets.

To interpret these values, we apply the heuristic proposed in
\autoref{sec:interpreting-uncert}. Since there are 30 authors in the
GCJ dataset and 81~authors in the GH dataset, we can compute the
thresholds $\frac{1}{30} \approx 0.03$ and $\frac{1}{81} \approx 0.01$,
respectively. As a result, Tigress attains practical $k$-uncertainty in
this experiment, since its uncertainty score reaches above 0.97 for the
GCJ dataset and 0.99 for the GH dataset.

Another interesting result of this experiment is that even with a
higher remaining accuracy, the uncertainty score for the code
normalization is better than for the imitation. While the imitation of
coding style more consistently causes misclassifications, the
confidence values often remain indicative of the authors. In contrast,
the normalization unifies the same stylistic patterns regardless of the
original author, thus creating a tighter anonymity set. In view of the
complex construction of adversarial examples for imitation, normalizing
the source code is a reasonable defense that preserves a good level of
readability and reduces stylistic patterns.

\medskip
\begin{tcolorbox}[boxrule=0.75pt,colback=white,colframe=cbone,sharp corners=all]
  \emph{Takeaway message.~} In the static attribution scenario, all
  techniques reduce the accuracy of the attribution methods. The
  achieved protection, however, varies between the techniques, with the
  obfuscator Tigress providing the best performance and achieving
  practical $k$-uncertainty when the adversary does not adapt to the
  anonymization.
\end{tcolorbox}

\subsection{Adaptive Attribution Scenario}
\label{sec:results-adapt-attr}

In our second scenario, we consider an \emph{adaptive attribution}, in
which the adversary is aware of the anonymization. As developing
countermeasures for each of the considered techniques is tedious, we
use a common trick from the area of adversarial machine learning: We
employ two simple variants of
\emph{adversarial \mbox{training}}~\citep{GooShlSze15} that enable the
learning algorithms to extract clues from the modified source code of
any possible anonymization strategy.

For code normalization and coding style imitation, we simply augment
the training data with modified samples. That is, we provide the
original source code and a normalized or imitated version of it. Since
both candidate techniques are easily overlooked by an attacker in
practice, this augmentation ensures that the attribution methods can
capture stylistic patterns from \emph{both}, the original and the
modified code. As a result, the methods are applicable regardless of
whether the candidate techniques are used or not. To also account for
this situation in the performance evaluation, we extend the test data
by providing both versions of the source code.

For obfuscation, we pursue a different variant of adversarial training.
In this case, the attacker can easily spot whether a source code has
been modified and hence we train the attribution methods on obfuscated
code only.  This strategy forces the attribution method to hunt for
subtle clues in the modified code, despite randomized names,
virtualized functions, and obscured control flow.

\begin{figure}[b]
  \centering
  \vspace{-4pt}
  \input{./pgf/acc_adaptive.pgf}
  \vspace{-18pt}
  \caption{Attribution performance (accuracy) of candidate techniques
    in the \emph{adaptive} attribution scenario (adversarial training).}
  \label{fig:retrained_acc}
\end{figure}

\subsubsection{Attribution Performance}

\autoref{fig:retrained_acc} presents the attribution performance for
the different techniques in the adaptive scenario on both datasets. A
notable drop in performance is not observable anymore. The accuracy of
all techniques remains over 50\% for the GCJ dataset and 30\% for the
GH dataset, except for Tigress. The obfuscator reduces the accuracy to
at most 25\% on the GCJ dataset and 8\% on the GH dataset. Still, the
remaining accuracy is significantly better than random guessing, which
would correspond to 3\% for the GCJ dataset and 1\% for the GH dataset.
Consequently, the attribution methods are capable of identifying some
developers despite strong obfuscation. While the real-world code in the
GH dataset increases the efficacy of Tigress, there is room for
improving the protection of the 8\% remaining developers.

Moreover, the impact of the adaptive attribution is particularly strong
for normalization and imitations of code. While for the static scenario
both techniques provide some protection, we now observe no defense and
in some cases even an improved performance. The weakness of the
techniques is that they aim to modify specific aspects of the source
code, but do not conduct broader transformations. These minor
modifications are compensated by the adversarial training, so that the
learning model can even better generalize the remaining coding style.

\subsubsection{Anonymization Performance}

The weak protection in the adaptive attribution is also reflected in
the uncertainty scores shown in \autoref{fig:retrained_uscore} for both
datasets. Compared to the static scenario, there is no clear tendency
for the values to increase. In several cases, the scores even decrease,
suggesting that the authors are now better identified than before
anonymization. Based on our heuristic, none of the approaches provides
adequate protection except for \emph{one} configuration. The obfuscator
Tigress achieves practical uncertainty for the method of \caliskan on
the GH dataset, as its uncertainty value is above 0.99. For \abuhamad's
method, however, only a value of 0.97 is achieved. Consequently, our
simple variants of adversarial training are already sufficient to
largely remove the protection of the candidate methods.

We observe another phenomenon: In several cases, the uncertainty score
increases for the method of \abuhamad while it decreases for the
approach of \caliskan To investigate this, we analyze the distribution
of uncertainty scores. The corresponding histograms for the GCJ dataset
are shown in \autoref{fig:u-scores-after}. The method of \caliskan
leads to a one-sided distribution. Between 40\% to 60\% of the authors
cannot be identified well. In contrast, the approach of \abuhamad
induces a two-sided distribution. Some authors are well protected while
others are perfectly identifiable. We attribute this observation to the
tendency of neural networks, as used by \abuhamad, to not generalize in
all cases.

While adversarial training does not completely eliminate the effect of
the four candidate techniques, it weakens the attained protection
considerably. Given that this approach is a simple countermeasure and
more advanced strategies can be conceived, we have to conclude that
\emph{none} of the techniques is a reliable solution for code
anonymization if an adversary is aware of their application. Still, we
note that strong obfuscation eliminates several clues from the code
despite adversarial learning and thus closing the remaining gap is a
promising direction for future research.

\smallskip
\begin{tcolorbox}[boxrule=0.75pt,colback=white,colframe=cbone,sharp corners=all]
  \emph{Takeaway message.~} In the adaptive attribution scenario, the
  attribution methods are weakly affected by the considered
  techniques, and the majority of authors remains identifiable.  The
  obfuscator Tigress provides the best protection, yet it fails to
  reach practical $k$-uncertainty in all cases, indicating a need
  for improved protection.
\end{tcolorbox}

\begin{figure}[t]
  \centering
  \vspace{-4pt}
  \input{./pgf/uncert_adaptive.pgf}
  \vspace{-18pt}
  \caption{Anonymization performance (uncertainty score) in the
    \emph{adaptive} attribution scenario (adversarial training).}
  \label{fig:retrained_uscore}
\end{figure}

\begin{figure}[b]
  \centering
  \vspace{-4pt}
  \scalebox{0.88}{
    \adjustbox{clip=true}{\input{./pgf/uncert_histo_caliskan.pgf}}
  }
  \scalebox{0.88}{
    \adjustbox{clip=true,trim=0 0 0 30}{\input{./pgf/uncert_histo_abuhamad.pgf}}
  }
  \vspace{-6pt}
  \caption{Distribution of uncertainty scores for the adaptive
    scenario on the GCJ dataset. Top: \caliskan Bottom: \abuhamad}
  \label{fig:u-scores-after}
\end{figure}

\subsection{Alternative Measures}
\label{sec:alternative-measures}

In \autoref{sec:uscore-vs-acc}, we argue that the proposed
uncertainty score considers different information than existing
performance measures and thus provides a better view on the
anonymization of developers. To investigate this claim, we conduct an
additional experiment in which we measure the \emph{correlation}
between the uncertainty score and alternative performance measures.

For this experiment, we consider the \emph{accuracy} as a traditional
measure of performance. Furthermore, we consider the
\emph{top-5 accuracy}, the \emph{mean reciprocal rank (MRR)} and the
\emph{normalized discounted cumulative gain (nDCG)} as three measures
from the field of information retrieval suitable for evaluating
rankings. We focus on the top 5 prediction, as our uncertainty score is
also calculated over the 5 neighboring confidence values. To determine
the correlation between these measures, we examine the attribution
results on the GCJ dataset. Since we apply 2~attribution methods,
2~scenarios, 4~candidate techniques and a grouped cross-validation over
8~source files, as well as the results of the unmodified source code,
we obtain a total of 144 performance values
($2 \times 2 \times 4 \times 8 + 2 \times 8$) for each measure. Based
on these values, we calculate the correlation coefficient for each pair
of the measures.

The results of this experiment are shown in \autoref{fig:corre}. A
clear trend can be seen in the resulting correlation matrix. The ranked
and unranked performance measures are more strongly correlated with
each other than with the uncertainty score. That is, the correlation
coefficients between the accuracy, the top-5 accuracy the, the MRR and
the nDCG range between 0.757 and 0.999. In contrast, the uncertainty
score correlates less with these measures. The highest values are
observed for the accuracy at 0.644. The measures derived from
information retrieval, which are attractive due to their ability to
take rankings into account, only achieve correlation coefficients
between 0.496 and 0.62.

These results are not sufficient to demonstrate that the proposed
uncertainty score is the only suitable measure for anonymization, yet
they show that existing approaches from machine learning and
information retrieval rely on other information and cannot be used as a
simple alternative. Following our reasoning from
\autoref{sec:uscore-vs-acc} on the role of an anonymity set in this
measurement, we therefore argue that the proposed uncertainty score is
better suited to describe how a developer is protected from
identification than classical performance measures.

\begin{figure}[t]
  \centering
  \scalebox{0.7}{\input{./pgf/corr_matrix.pgf}}
  \vspace{-10pt}
  \caption{Correlation of alternative performance measures with the
    uncertainty score.}
  \label{fig:corre}
\end{figure}

\section{Anonymization Deficits}
\label{sec:impr-code-anonym}

Our empirical analysis demonstrates that the four candidate techniques
offer only limited protection in practice. In this section, we take a
closer look on this problem and introduce two methods for explaining
the decisions of attribution. Based on these explanations, we then
uncover clues left by the techniques in the source code. This analysis
enables us to finally improve Tigress, as the best approach in our
experiments, and iteratively remove remaining clues.

\subsection{Understanding Attribution}
\label{subsec:impr-code-anonym-understanding}

We introduce two strategies to understand why an attribution is still
possible after a candidate technique has been applied to a program:
\emph{feature highlighting} and \emph{occlusion analysis}.

\subsubsection{Feature Highlighting}

A simple yet effective way to explain an attribution is to trace back
the decision of the learning algorithm to individual features of the
code. To this end, we adjust the feature extractions to collect the
\emph{code regions} associated with each feature. For AST-based
features, these regions can be easily determined using the Clang
frontend. Only a few features, such as the depth of the AST, have no
specific code region and are thus omitted.

Based on this mapping, we apply \emph{explanation methods} to trace
back the attributions to code regions \citep{WarArpWreRie+20}. For
example, for the random forest classifier employed by \caliskan, we use
the method \emph{TreeInterpreter}~\cite{treeinterpreter}, which returns
the contribution of every tree node to the prediction. We then color
the code regions based on this relevance.
\autoref{fig:impr-code-anonym-feat-highlighting} exemplifies the
explanation for a code snippet after applying Stunnix. The includes,
declarations, and API usages are shaded in darker color, indicating
that they still provide clues for authorship attribution. In fact,
these patterns consistently occur for the respective author in our
evaluation.

\begin{figure}[bp]
  \centering
  \fbox{
    \input{./tikz/impr-code-anonym-feat-highlighting.tex}%
    \hspace{4.5cm}
  }
  \caption{Example of feature highlighting for explaining an
    attribution. Darker shading indicates more relevance.}
  \label{fig:impr-code-anonym-feat-highlighting}
\end{figure}

\subsubsection{Occlusion Analysis}

Feature highlighting is particularly effective for explaining the
attribution of mildly modified code. For strong obfuscation, however,
it reaches its limits. While we can highlight areas in the obfuscated
code generated by Tigress, these are incomprehensible by design and
impede further analysis. To address this problem, we introduce
\emph{occlusion analysis}. Similarly to the field of computer vision,
where classifications are often explained by occluding regions of an
image~\citep{ZeiFer14}, we occlude areas of the source code and observe
the resulting attribution.

\autoref{alg:occlus} provides an overview of this approach. First, we
partition the unobfuscated code into segments $S$~(line~2). Then, we
iteratively remove each segment $s \in S$ from the code, apply the
obfuscation~$\anonym$ and perform the attribution~$\attr$
\mbox{(lines~4--8)}. After this step, the relevance $R_s$ of the
segment $s$ is given by the confidence difference to the original
attribution (line 7). We repeat this process, so that we obtain a
relevance map over all segments.

\begin{algorithm}[tp]
  \caption{Explaining attributions with occlusions}
  \label{alg:occlus} \small
  \vspace{3pt}
  \begin{algorithmic}[1]

    \Require Program $\prg$, attribution $\attr$,
    anonymization $\anonym$, author~$t$

    \State $c_t^* \leftarrow \attr(\anonym(\prg))$

    \State $S \leftarrow \Call{segmentCode}{\prg}$
    \Comment{Line splitting / program slicing}
    \State $R \leftarrow (0, \ldots, 0) \in \mathbb{R}^{|S|}$
    \Comment{Initialize relevance vector}

    \ForAll {$s \in S$}
    \State $\prg^s \leftarrow \Call{occludeSegment}{\prg, s} $
    \State $c^s \leftarrow \attr(\anonym(\prg^s))$
    \Comment{Attribution w/o segment $s$}
    \State $R_s \leftarrow (c_t^*  - c_t^s)$
    \Comment{Relevance of segment $s$}
    \EndFor

  \end{algorithmic}
\end{algorithm}

While the method can be applied to every anonymization techniques, we
cannot remove arbitrary parts of a program without affecting its
syntax. To address this problem, we introduce two strategies: As the
first strategy, we split the source code along the textual lines. This
approach naturally leads to incorrect syntax, yet often the remaining
code is still valid and we can narrow down relevant code lines. As the
second strategy, we employ \emph{backward program slicing}. That is, we
use the framework \mbox{\emph{Frama-C}}~\citep{CuoFloPrev+12} which
enables creating syntactically correct program slices on C code. This
strategy preserves the syntax, yet the segments often become large,
making an identification of relevant regions difficult.

\subsection{Identified Code Clues}
\label{subsec:impr-code-anonym-identified-clues}

With the help of the explanation methods, we investigate the deficits
of the candidate techniques on the GCJ dataset. After manually
analyzing the highlighted code regions with both strategies, we
identify four recurring groups of patterns that remain in the code.

\subsubsection{String Literals}

The first group of patterns corresponds to string literals. Code
normalization and coding style imitation do not modify these, Stunnix
just replaces strings with their hexadecimal representations.
Therefore, a learning algorithm can use them to find clues about the
developers. In contrast, Tigress takes care to not reveal literals by
dynamically generating strings at runtime. While the characters
themselves are not present, the code necessary for their generation
still leaves telltale signs, \eg~the length of the strings is
implicitly reflected in the size of the generation routine. As a
result, some subtle hints remain in the obfuscated code.

\subsubsection{Include Directives}

Another group is formed by \code{\#include} directives, which reveal a
developer's preferences for certain functions and libraries. Neither
code normalization nor the obfuscator Stunnix touch these directives
and thus expose these patterns to the attribution methods, as also
highlighted in \autoref{fig:impr-code-anonym-feat-highlighting}. The
coding style imitation by \mbox{\citet{QuiMaiRie19}} inserts and
removes include directives to match other developers. Nevertheless,
headers required for the implementation always remain in the code.
Finally, Tigress ``inlines'' the headers by copying their content into
the source code. While this makes the resulting code hard to understand
for a human, the included content is no different from the directive
for a learning algorithm and thus still serves as a valuable hint.

\subsubsection{API Usage}

API usage provides another set of patterns that remains after
anonymization. Automatically changing it is a challenging task, since
one must ensure that the replacement is equivalent in functionality.
Although the coding style imitation contains some transformations to
exchange equivalent C functions, the majority of API calls remains
unchanged. The code normalization and Stunnix provide no
transformations for this. Tigress calls the API functions by their
addresses, so that it can hide function names. Nevertheless, an
attribution method can use the types and number of call parameters to
narrow down the particular API. In \appref{sec:appendix-api-tigress},
we provide a more detailed analysis of this remaining feature.

\subsubsection{Code Structure}

Finally, the program structure is often preserved. With the exception
of Tigress, the other techniques retain the general organization of the
source code. Although the coding style imitation is able to rearrange C
statements locally, the overall structure stays unchanged. As a result,
personal preferences to structure the program are available to the
attribution methods.

\subsection{Eliminating Code Clues}
\label{subsec:impr-code-anonym-eliminate-clues}

Equipped with knowledge of indicative patterns in the modified code, we
are ready to refine the anonymization of source code. For this
improvement, we focus on the obfuscator Tigress, as it provides the
best protection in our experiments.

\subsubsection{Code Transformations}

To eliminate the identified patterns, we design a set of code
transformations that addresses the weak spots. We iteratively apply new
transformations and then observe their impact on the attribution using
the methods from \autoref{subsec:impr-code-anonym-understanding}. This
feedback loop enables us to \emph{systematically} identify and
eliminate clues left in the code, increasing the attained
$k$-uncertainty.

In particular, we devise the following transformations: To hide string
literals, we remove empty function stubs inserted by Tigress and pad
all strings to a minimum length. Furthermore, we include all headers
from the C standard by default and add at least one call to every API
function used in the dataset. For function pointers, we remove all
information, except for necessary argument and return types. This makes
it complicated to identify the called functions.

\begin{table}[t]
  \caption{Performance of Tigress with eliminated clues in the
    adaptive attribution scenario on the GCJ dataset. The numbers in
    brackets show the difference to the results on unmodified code.}
  \label{tab:improved}
  \centering \small
  \begin{tabular}{lcc}
    \toprule
    \bf Attribution & \bf Accuracy                      & \bf Uncertainty score            \\
    \midrule
    \caliskan       & 0.071 ~\textcolor{old}{(--0.617)} & 0.969 ~\textcolor{old}{(+0.128)} \\
    \abuhamad       & 0.058 ~\textcolor{old}{(--0.696)} & 0.938 ~\textcolor{old}{(+0.677)} \\
    \bottomrule
  \end{tabular}
\end{table}

\subsubsection{Results}

\autoref{tab:improved} shows the attribution performance after applying
these improvements and conducting another run of adversarial training
on the GCJ dataset. We observe a significant decrease in accuracy
compared to \autoref{tab:default} and \autoref{fig:retrained_acc}. The
values are close to guessing and the uncertainty scores are comparable
to the static scenario (see \autoref{tab:default}). For the method of
\caliskan, the score reaches the threshold of 0.96, so that we attain
practical $k$-uncertainty also for the GCJ dataset. For the method of
\abuhamad, we come close to this with an uncertainty value of 0.94, but
a minor gap still remains.

This positive outcome may seem like the final defeat of the two
attribution methods. Unfortunately, this is a misinterpretation of the
conducted experiments. We only show that it is possible to achieve
$k$-uncertainty in a controlled environment where the defender can
systematically explore the attacker's capabilities for attribution. In
practice, however, this is rarely the case, and so we demonstrate the
technical feasibility of attaining $k$-uncertainty in an adaptive
scenario but unfortunately not its general realization.

\smallskip
\begin{tcolorbox}[boxrule=0.75pt,colback=white,colframe=cbone,sharp corners=all]
  \emph{Takeaway message.~} It is possible to attain
  \mbox{$k$-uncertainty} by systematically identifying and eliminating
  indicative clues in source code. However, this approach is only
  tractable if the defender operates in a controlled setup and has
  access to the attribution method, which is rarely the case in
  practice.
\end{tcolorbox}

\section{Limitations}
\label{sec:disc--limit}

With our theoretical and practical analysis, we shed light on
challenges of anonymizing code. Naturally, our approach to tackle this
problem comes with limitations that we discuss in the following.

\subsection{Selection of Techniques}

For our experiments, we select two attribution methods and four
anonymization techniques. Consequently, our results are based on this
particular choice. Nonetheless, we argue that the obtained results are
unlikely to be completely different for other selections due to the
following reasons:

\medskip
\emph{Attribution Methods.}
We consider two state-of-the-art methods for authorship attribution.
While other approaches would also be applicable
\citep[e.g.,][]{BogKovRebBac2021, DauCalHarShe2019, DinFunIqbChe2019},
none of these is fundamentally different in design. All methods extract
layout, lexical, and syntactic features. Since the two considered
methods already substantially weaken the anonymization, evaluating more
attribution methods without new strategies would not provide further
insights.

\medskip
\emph{Anonymization Techniques.}
For code anonymization, we consider common approaches for reducing the
impact of coding style. With Tigress, we employ one of the most
powerful obfuscators available for C code
\mbox{\citep{BanColPre17, tigress, NagCol09}}. Our analysis in
\autoref{sec:impr-code-anonym} demonstrates how this obfuscation can be
improved to realize $k$-uncertainty in a controlled setting. The four
selected techniques thus provide a broad view on current defenses
against authorship attribution. However, we concede that more advanced
anonymization strategies are clearly conceivable and hope to encourage
more work in this direction with our theoretical and practical analysis.

As a promising candidate for further improvements, we have started
experimenting with large language models. These models allow the
generation and transformation of source code in different contexts.
Due to their probabilistic sampling process, they may serve as an
alternative approach to manipulating and obfuscating coding style. In a
first experiment, we therefore instructed the OpenAI model GPT-3.5 to
change the coding style of the source files from our GCJ dataset. While
the model returned modified code for all examples, we found that 11\%
of them were syntactically incorrect and another 37\% unfortunately had
different semantics or crashed. For this reason, we have not taken any
further steps here and will leave this exploration to future work.

\subsection{Size and Type of Code}

Compared to prior work, we focus on two small datasets with only 30 and
81 authors, respectively. The reason for this limitation is that we
restrict our experiments to plain C code, since advanced
transformations on C++ are challenging and not supported by Tigress.
Previous work has demonstrated that the performance of learning-based
attributions methods decreases gradually with the number of considered
authors~\mbox{\citep[see][]{IslHarLiuNar+15, AbuAbuMohNya+18}}. Hence,
the lack of protection observed in our experiments may disappear once
the attacker has to consider a large set of authors. However, this set
is chosen by the attacker during training and cannot be controlled by
the defender directly. A reliable protection should thus be also
effective for a small number of authors.

Finally, we focus on C code because it is widely used in software
development. Still, we note that interpreted languages, such as Python
and JavaScript, offer further strategies for anonymization, including
encrypting code and unpacking it at runtime. Although there exists a
large series of research on unpacking malicious
code~\citep{UgaBalSanBri+15, KolLivZorSei+12} that would reveal the
original code, investigating other types of protecting
code---compilation vs. interpretation---may provide further strategies
for improving protection in practice.

\subsection{Undecidability}

The concept of $k$-uncertainty inherits the \emph{undecidability} from
$k$-anonymity. It is unfortunately impossible to create an
anonymization method that can guarantee $k$-uncertainty for any
possible attribution method and value of $\epsilon$. We argue, however,
that $k$-uncertainty provides an advantage over $k$-anonymity: It
involves a tunable confidence range $\epsilon$. By making this a
measurable quantity, we create the uncertainty score that allows us to
compare existing methods, which would not be possible with
$k$-anonymity.

While guaranteed anonymity of source code would be preferable and might
be attainable in controlled environments, our main result is negative.
Nonetheless, we believe that this negative outcome is a central insight
that advances research on protecting developers in practice by shaping
directions for future research.

\subsection{Confidence Values}

The uncertainty score builds on the concept of confidence to assess how
well a developer is protected. Unfortunately, not all learning
algorithms implement this concept to the same extent. While some
algorithms return proper confidence values, others only provide output
normalized to a range between 0 and 1. In these cases, the uncertainty
score only measures the relative proximity of the authors, without an
appropriate interpretation of confidence. Similarly, our heuristic for
determining practical uncertainty only provides a good estimate if
suitable confidence values are provided.

In our experiments, we employ the confidence values returned by a
random forest, which correspond to the mean predicted class
probabilities of the trees in the forest. These values are based on a
reasonable notion of confidence and therefore do not invalidate our
analysis. However, when conducting experiments with other learning
models, we recommend carefully examining the concept of confidence to
avoid misinterpreting the uncertainty score.

\section{Related Work}
\label{sec:related-work}

Our work is the first to explore the problem of anonymizing source
code. However, we naturally build on previous research from different
areas, such as code authorship attribution and data anonymization. In
the following, we briefly discuss these related branches.

\subsection{Code Authorship Attribution}

The starting point for our work has been the remarkable progress in
code stylometry, that is, the authorship attribution of code.

\subsubsection{Code Stylometry}

Several methods have been developed that are able to almost perfectly
attribute single-author code to developers using different concepts of
machine learning~\mbox{\citep[e.g.,][]{IslHarLiuNar+15,
        AbuAbuMohNya+18, AlsDauHarMan+17, WanJiWan2018}}. These methods
have been further extended to attribute code fragments written by
multiple authors~\mbox{\citep[e.g.,][]{DauCalHarShe2019,
        TerPasBlaSua2022, BogKovRebBac2021}}. This partial attribution,
however, proves challenging and therefore leads to lower detection
rates. For this reason, we focus our empirical analysis on methods
analyzing single-author code. In this way, we evaluate techniques for
protecting code under a stronger adversary model.

\subsubsection{Coding Style Imitation}

Another branch of research has explored the robustness of
learning-based attribution methods. In the first study by
\citet{SimZetKoh2018}, manual modifications have been used to mimic the
style of developers. Following work has then developed concepts for
automatically creating adversarial examples of source
code~\cite{QuiMaiRie19, LiuJiLiuWu2021, MatStaPrePer+19}. These attacks
differ in the employed code transformations and search strategy. For
example, \citet{QuiMaiRie19} and \citet{LiuJiLiuWu2021} develop several
code transformations that modify lexical and syntactic features,
whereas \mbox{\citet{MatStaPrePer+19}} apply rather simple
modifications such as changing the layout or copying comments. For our
evaluation, we focus on the attack by \mbox{\citet{QuiMaiRie19}},
as it has a higher evasion rate than the method of
\citet{LiuJiLiuWu2021} and allows changing various lexical and
syntactic features in C code.

\subsubsection{Text Stylometry}

Finally, there is extensive work on attributing authorship of natural
language texts and imitating the style of writing. Examples of this
research include techniques for detecting patterns in writing
style~\mbox{\citep[e.g.,][]{StoOveAfrGre2014, DinFunIqbChe2019,
        AfrIslStoGre+14}} as well as approaches for misleading an
attribution through writing style obfuscation
\citep[e.g.,][]{BreAfrGre2012, MahAhmShaSri2019, McDAfrCalSto2012}. Our
work shares inspiration from this. Due to the fundamentally different
properties of natural language and source code, however, these
approaches are not directly applicable in our setting.

\subsection{Data Anonymization}

Another related area is the anonymization and de-anonymization of
data~\citep[e.g.,][]{NarShm08, MacKifGehVen07, HaePieNar11, Swe02}.
Early ideas of this area originate from general data processing and
tackle the challenges of analyzing and exchanging privacy-sensitive
data, such as medical records.

\subsubsection{Anonymity Concepts}

One of the first ideas from this area is the concept of
\mbox{\textit{k-anonymity}} by \citet{Swe02}. In a dataset, every
quasi-identifier needs to be hidden in a group of at least $k$ persons
with the same identifier, called the anonymity set. This set ensures
that no individual can be isolated through personal properties.

The concept of $k$-anonymity, however, is insufficient when additional
data is correlated with the anonymity set. This has led to the
development of \textit{$\ell$-diversity}~\citep{MacKifGehVen07}. This
concept requires for every group of equal quasi-identifiers that at
least $\ell$ different sensitive attributes are also included. In this
case, even if an individual can be assigned to a certain equivalence
class, the attacker is not able to deduce further sensitive data. This
concept was further improved by \textit{t-closeness}~\citep{LiLiVen07},
which tackles the problem of information disclosure through the
different distributions of attributes in an equivalence class and the
overall data. This concept requires a similar distribution in both.
Hence, an attacker is not able to learn more about a specific
individual than about the dataset.

Unfortunately, we conclude from our theoretical analysis that
\mbox{$\ell$-diversity} and \mbox{$t$-closeness} are not helpful for
protecting code, since $k$-anonymity is incomputable for an unknown
attribution method.

\subsubsection{Differential Privacy}

Finally, our work also relates to the powerful concept of
\textit{differential privacy}~\citep{Dwo06}. In this concept,
privacy-sensitive data is not directly available to users but provided
through an interface (or post-processing step). By adding carefully
chosen noise to the answers of this interface, it becomes impossible to
tell whether an individual is present in the data or not. This concept
has recently gained popularity as a strategy for improving the privacy
of data in learning models~\citep[e.g.,][]{ChaMonSar11, SuCaoLiBer+16,
    AbaChuGooMcM+16, HeMacFlySri+17, JayEva19} and also in the field of
natural language processing~\mbox{\citep[e.g.,][]{WegKer18, LyuHeLi20,
        FleRoeHud21, MatWegKer22}}.

In natural language processing, the noise is usually not added to the
text itself, but to a vector representation of it
\mbox{\citep{WegKer18, LyuHeLi20, FleRoeHud21}}. While this is an
elegant approach, in our setting, this requires knowledge and access to
the feature representation used by the attacker. As this is typically
unknown to the defender, these approaches are not directly applicable
to protect developers from identification.

\section{Conclusion}
\label{sec:conclusion}

Methods for authorship attribution of source code have substantially
improved in recent years. While first approaches have suffered from low
accuracy, recent techniques can precisely pinpoint a single developer
among hundreds of others. Defenses against this progress have received
little attention so far and hence we provide the first analysis of code
anonymization. Theoretically, we reveal a strong asymmetry between
attackers and defenders, where the universal $k$-anonymity of programs
is generally undecidable. Practically, however, we provide a framework
for reasoning about and measuring anonymity using the concept of
$k$-uncertainty.

Although we can generate $k$-uncertainty in a controlled setup, the
main conclusion of our empirical analysis is negative: We find that
effective techniques for protecting the identity of developers in
practice are still lacking. Research on such techniques is challenging,
as the defender is naturally not aware of all possible strategies for
attribution, while the attacker can easily compensate new anonymization
methods through adversarial training, as we demonstrate in our
experiments.

In summary, we conclude that entirely new approaches to anonymization
are needed, possibly starting already in program language design and
software development. For example, new program languages and
environments could be designed with anonymity in mind, so that
stylistic patterns and telltale clues are reduced during development,
potentially creating a unified mapping between semantically equivalent
code and its representation. Our work is a first step in this direction
and provides concepts for defining and measuring code anonymity in such
future settings.

\section*{Public Code and Data}
\label{sec:public-code-data}

To encourage further research on code anonymity, we make the
implementation of our methods publicly available. In addition, we
provide the collected source code so that all experiments can be
reproduced and serve as basis for evaluating new approaches.
\begin{center}
  \color{primarycolor!90}\tt\small
  \url{https://github.com/horlabs/anonymizer}
\end{center}

\begin{acks}
  This work was funded by the German Federal Ministry of Education and
  Research in the project IVAN (16KIS1165K) and under the grant
  BIFOLD23B, the Deutsche Forschungsgemeinschaft (DFG, German Research
  Foundation) under Germany’s Excellence Strategy – EXC 2092 CASA –
  (390781972), and the European Research Council (ERC) under the
  consolidator grant MALFOY (101043410). Moreover, this work was
  supported by the IFI program of the German Academic Exchange Service
  (DAAD) funded by the Federal Ministry of Education and Research
  (BMBF).
\end{acks}

\bibliographystyle{abbrvnatx}
\bibliography{anonymizer}

\appendix

\section{Examples of Equivalent Code}

\autoref{fig:examples} shows four implementations of the Euclidean
algorithm in C. The programs are semantically equivalent but make use
of different identifiers, types, arithmetics, and control flow. They
serve as a simplified example for the variety of coding styles and
possible program representations.

\vspace{-2mm}
\begin{figure}[htbp]
  \subfloat[Variant 1]{
    \begin{minipage}{0.22\textwidth}
      \scriptsize
\begin{minted}{c}
 /* Vanilla version */
 
 int gcd(int a, int b) {
 
    while (b != 0) {
       int tmp = b;
       b = a % b;
       a = tmp;
    }

    return a;
 }
\end{minted}
    \end{minipage}}
  ~~
  \subfloat[Variant 2]{
    \begin{minipage}{0.22\textwidth}
      \scriptsize
\begin{minted}{c}
 typedef int num;

 num gcd(num n, num m) {
  loop:
    if (n > m)
       n = n - m;
    else if (n < m)
       m = m - n;
    else
       return n;
    goto loop;
 }
\end{minted}
    \end{minipage}}

  \subfloat[Variant 3]{
    \begin{minipage}{0.22\textwidth}
      \scriptsize
\begin{minted}{c}
#include <stdint.h>

int32_t gcd(int32_t x,
	    int32_t y) {
   if (y == 0)
      return x;

   int32_t z =
       x - x / y * y;

   return gcd(y, z);
}
\end{minted}
    \end{minipage}}
  ~~
  \subfloat[Variant 4]{
    \begin{minipage}{0.22\textwidth}
      \scriptsize
\begin{minted}{c}
 #include <math.h>
 
 int gcd(int a, int b) {
    long c;

    for(c = 0; b > 0; 
        c = !(a = c)) {
        c = c + b;
        b = fmodl(a,b);
    }
    return a;
 }
\end{minted}
    \end{minipage}}

  \caption{Semantically equivalent programs implementing the Euclidean
    algorithm.}
  \label{fig:examples}
\end{figure}

\section{GitHub Dataset}
\label{sec:gith-crawl-proc}

For our experiments, we assemble a dataset of source code from GitHub.
Our collection procedure consists of three steps, where we first crawl
repositories with source code (\autoref{sec:crawling-step}), then
filter them to match our experimental setup
(\autoref{sec:filtering-step}), and finally configure proper build
environments (\autoref{sec:configuration-step}).

\subsection{Crawling Step}
\label{sec:crawling-step}

To obtain a wide range of repositories, we implement a recursive crawl
of Github. The crawl starts with the curated software list
\emph{awesome-c}\footnote{\url{https://github.com/oz123/awesome-c}},
which contains a collection of popular software projects written in C
and hosted on Github. From this list, we retrieve metadata from all
repositories and extract developers who have contributed to them. We
then recursively visit the repositories of these developers. To avoid
an explosion of the recursion, we limit the crawling using the
following criteria:
\begin{enumerate}
  \item We ignore all repositories not marked as C code.
  \item We ignore all forked repositories to avoid duplicates.
  \item We ignore large repositories with over 100 authors.
  \item We ignore large repositories with over 10\,Mb codebase.
\end{enumerate}
We terminate the crawl after a period of 24 hours. While this strategy
results in a large list of potential source code, most of the retrieved
repositories do not satisfy the requirements of our experimental setup.
For instance, the majority of C~projects are developed collaboratively,
so source files can rarely be attributed to individual authors. This
mix of authorship poses a problem for attribution, and we refer the
reader to the work of \mbox{\citet{DauCalHarShe2019}} for a
corresponding discussion.

\subsection{Filtering Step}
\label{sec:filtering-step}

Next, we employ a filtering step to remove repositories and source code
unsuitable for our experimental setup. In particular, this step filters
the repositories based on the following criteria:
\begin{enumerate}

  \item We remove all source files with less than 50 lines of code, as
        they are too short for inferring coding style.

  \item We remove all source files with more than 5,000 lines of code,
        as many of these contain large chunks of constant data or
        automatically generated code.

  \item We remove all source files whose commit messages contain the
        words ``signed off'' or ``copied'', which indicates that the commit user may not be the author.

  \item We remove all source files with less than 5 commits. Several
        projects copy code from other repositories, which is indicated by a lack of active development.

  \item We remove all source files where less than 90\% of the lines of
        code are not developed by a single author. We allow a gap of 10\% to compensate for editorial changes.

\end{enumerate}

After the data has been filtered at file level, we remove all empty
repositories and the corresponding authors. Finally, to allow splitting
the data into training and test partitions, as described in
\autoref{sec:evaluation-setup}, we keep only those authors who
contributed to at least 2 repositories, so that we can use one
repository for testing and the others for training. We also require
that the training repositories contain at least 7 files. This ensures
that the split into training and testing is similar to the GCJ dataset
and therefore the performance results are comparable.

\subsection{Configuration Step}
\label{sec:configuration-step}

In the last step, we try to configure the build environment for each
repository. For this purpose, we execute supplied configuration
scripts, such as \emph{cmake} and \emph{configure}. To obtain a large
number of correctly configured projects, we manually install
dependencies and set configuration options where possible. We then test
the resulting code against the anonymization techniques under test and
remove those files that cannot be processed correctly, for example,
because they require complicated dependencies or use unusual
programming features. Overall, this step is the most time-consuming and
takes about a whole person-month. As a result, the available source
code for our experiments is drastically reduced. In the end, we have
81~authors, 391~repositories and 1,284~files of source code that
successfully passes all stages of our experimental setup.

While we aim to provide a more realistic picture of source code than
the common GCJ dataset, we have to acknowledge that our restrictive
filtering limits its representativeness. However, we consider this
filtering a necessary compromise: While we could select a larger
proportion from the crawl for our evaluation, we would risk its ground
truth being wrong due to mislabeled authors and corrupted source code.
These defects could invalidate our experiments, which depend on
accurate analysis of identified developers. Therefore, we strive for a
balance between valid ground truth and representativeness by starting
from a large crawl and then successively removing error sources in the
underlying data.

\section{API Hiding of Tigress}
\label{sec:appendix-api-tigress}

Tigress hides the usage of an API by determining the addresses of the
API functions at runtime. Still, the types and number of parameters are
specified, because the function pointers must be typed according to the
passed parameters for every call. This is achieved by casting the
pointers using function declarations from the header files. Some
declarations include argument names and thus these are copied into the
corresponding casts. This makes it possible to differentiate functions
with the same types of parameters.

As an example, the functions \code{abs} and \code{close} require a
single parameter of type \code{int} and return the same data type. This
leads to a function pointer of type \code{int (*)(int)}. In the header
files, however, the parameter of \code{abs} is named \code{\_\_x},
while for \code{close} it is \code{filedes}. The corresponding casts in
the obfuscated file are therefore \codeintext{(int (*)(int \_\_x))} and
\codeintext{(int (*)(int filedes))} and thus are easily
distinguishable. As a result, even for Tigress, an attribution method
can identify used library functions in this case.

\newpage

\section{Normalization Rules}
\label{sec:appendix-normalization}

\autoref{tab:norm-rules} provides a detailed listing of the implemented
normalization rules for anonymization.

\begin{table}[h!]
  \renewcommand{\arraystretch}{1.4}
  \caption{Overview of implemented normalization rules}
  \label{tab:norm-rules}
  \begin{tabular}{>{\small\tt\color{cbone}}l>{\small}p{.55\columnwidth}}
    \toprule
    \bf \textcolor{black}{Rule name} & \bf Description                                              \\
    \midrule
    Renaming                         & All variables, functions, and structures are renamed to a
    generic version. For example, all variables are numbered
    as \code{var\_x} with \code{x} being a number
    starting at 0.                                                                                  \\
    Types                            & The used data types are mapped to a specified subset to
    eliminate redundant type names, such as \code{long} and
    \code{int\_32}. Note that this transformation is
    platform-specific.                                                                              \\
    Switch2If                        & \codeintext{switch} statements are transformed to a chain of
    \code{if-else} statements.                                                                      \\
    Comma                            & Comma operators are largely eliminated and replaced with a
    sequence of statements containing the expressions.                                              \\
    CompoundAssign                   & Compound assignments are replaced with normal
    assignments and the specified binary operator,
    e.~g. \code{a += 2} is transformed into
    \code{a = a + 2}.                                                                               \\
    IfElse                           & If the last statement inside the body of an \code{if}
    statement is for example a \code{return} or \code{break},
    the following code is moved into an \code{else} to this
    \code{if}.                                                                                      \\
    MainParams                       & This rule enforces the use of two parameters for the
    \code{main} function and a \code{return} statement at
    its end.                                                                                        \\
    Multidecl                        & If multiple declarations are in a single statement, the
    statement is split into separate declaration
    statements.                                                                                     \\
    Braces                           & Braces around every body are enforced, for example, for the
    bodies of all \code{if} and \code{for}
    statements.                                                                                     \\
    UnnecessaryReturn                & This rule removes \verb|return| statements in
    \code{if} bodies if all following code is in the
    \code{else} clause and the function has no
    return value.                                                                                   \\
    VoidReturn                       & This rule adds a \code{return} statement at the end of
    every \code{void} function.                                                                     \\
    FlattenIf                        & For nested \code{if} statements, this rule removes inner
    clauses by combining the conditions of the inner and
    outer \code{if}. It inserts an additional \code{if} for
    every \code{else}.                                                                              \\
    Paren                            & This rule removes unnecessary parentheses like in
    \code{a = (b + c)}, simplifying arithmetic expressions                                          \\
    \bottomrule
  \end{tabular}
\end{table}

\section{Used Transformations for Tigress}
\label{sec:appendix-tigress-transformations}

\autoref{tab:tigress-transformations} lists the used transformations and
arguments for obfuscation with Tigress in detail.

\begin{table}[h]
  \renewcommand{\arraystretch}{1.4}
  \caption{Overview of used transformations and arguments for source
    code obfuscation with Tigress}
  \label{tab:tigress-transformations}
  \begin{tabular}{>{\small\tt\color{cbone}}l>{\small}p{.55\columnwidth}}
    \toprule
    \bf \textcolor{black}{Transformation} & \bf Arguments \\
    \midrule
    InitEncodeExternal                    &
    Functions=main\newline
    InitEncodeExternalSymbols=\newline
    \hspace*{20pt}\verb|<ext. functions>|                 \\
    InitEntropy                           &
    InitEntropyKinds=vars \newline
    Functions=init\_tigress                               \\
    InitOpaque                            &
    Functions=init\_tigress \newline
    InitOpaqueStructs=env                                 \\
    RandomFuns                            &
    RandomFunsName=SECRET \newline
    RandomFunsFunctionCount=3 \newline
    RandomFunsCodeSize=20 \newline
    RandomFunsLoopSize=5                                  \\
    EncodeLiterals                        &
    Functions=\newline
    \hspace*{20pt}\verb|<all in file>|,main\_0,\newline
    \hspace*{20pt}/SECRET.*/ \newline
    EncodeLiteralsKinds=string \newline
    EncodeLiteralsEncoderName=\newline
    \hspace*{20pt}stringEncoder                           \\
    Merge                                 &
    MergeFlatten=false \newline
    MergeName=MERGED \newline
    Functions=\newline
    \hspace*{20pt}\verb|<w/o main>|,main\_0,\newline
    \hspace*{20pt}/SECRET.*/                              \\
    Virtualize                            &
    VirtualizeDispatch=switch \newline
    VirtualizeStackSize=48 \newline
    VirtualizeOperands=mixed \newline
    VirtualizeMaxDuplicateOps=2 \newline
    VirtualizeSuperOpsRatio=0.1 \newline
    VirtualizeMaxMergeLength=3 \newline
    Functions=MERGED,stringEncoder                        \\
    EncodeLiterals                        &
    Functions=\newline
    \hspace*{20pt}main,MERGED,stringEncoder \newline
    EncodeLiteralsKinds=integer                           \\
    EncodeExternal                        &
    Functions=MERGED \newline
    EncodeExternalSymbols=\newline
    \hspace*{20pt}\verb|<ext. functions>|                 \\
    CleanUp                               &
    CleanUpKinds=*                                        \\
    \bottomrule
  \end{tabular}
\end{table}

\end{document}